\documentclass[%
 reprint,
 amsmath,amssymb,
 aps,
]{revtex4-1}

\usepackage{graphicx}
\usepackage{dcolumn}
\usepackage{bm}

\usepackage{hyperref}
\begin{document}

\preprint{APS/123-QED}
\title{On the magnetization process in ferromagnetic materials}

\author{Ruben Khachaturyan}
 \email{rubenftf@gmail.com}
\author{Vahram Mekhitarian}%
 \email{vahram.mekhitarian@gmail.com}
\affiliation{%
 Institute for Physical Research, NAS of Armenia, Ashtarak, Armenia
}%
\date{\today}

\begin{abstract}
The present article concludes that a ferromagnetic sample could be considered as a paramagnetic system where roles of magnetic moments play magnetic domains. Based on this conclusion and taking into account presence of an anisotropy field the formula which describes magnetization dependence on the external magnetic field is derived. Expressions for a remanent magnetization and a coercive force are presented. The new parameter to characterize a magnetic stiffness of a material is introduced. A physical expression for a dynamic magnetic susceptibility as a function of material’s characteristics, external magnetic field, and temperature is given.
\end{abstract}

\pacs{Valid PACS appear here}
\maketitle


\section{\label{sec:level1}INTRODUCTION}

A physical theory permits correctly involve all interactions in a magnetization process and to reveal relationship between structure and physical properties of a magnetic material.

Applicable mathematical model could be derived from such a theory. This model will give a possibility to investigate real physical and structural properties of magnetic materials from experimental data. It is essential for synthesis of new materials with desired properties.

Nowadays there are several models for describing magnetization of ferromagnetic materials. More detailed description and analyzes of advantages and disadvantages of these models one can find in works \cite{1}-\cite{4}.

In the present paper, there suggests a new theory of magnetization and an attempt to derive an applicable general mathematical model to describe magnetization curve for soft and stiff magnetic materials. Such generic special points in magnetization curve are also elicited from the model.

A formula for dynamic magnetic susceptibility is derived from a magnetization – field dependence. 

A new parameter which can numerically characterizes the magnetic stiffness of a material is introduced and its physical interpretation is given.
\section{\label{sec:level1}MAIN IDEA}
\begin{figure}[h!]
	\centering
	\includegraphics[width=0.4\textwidth]{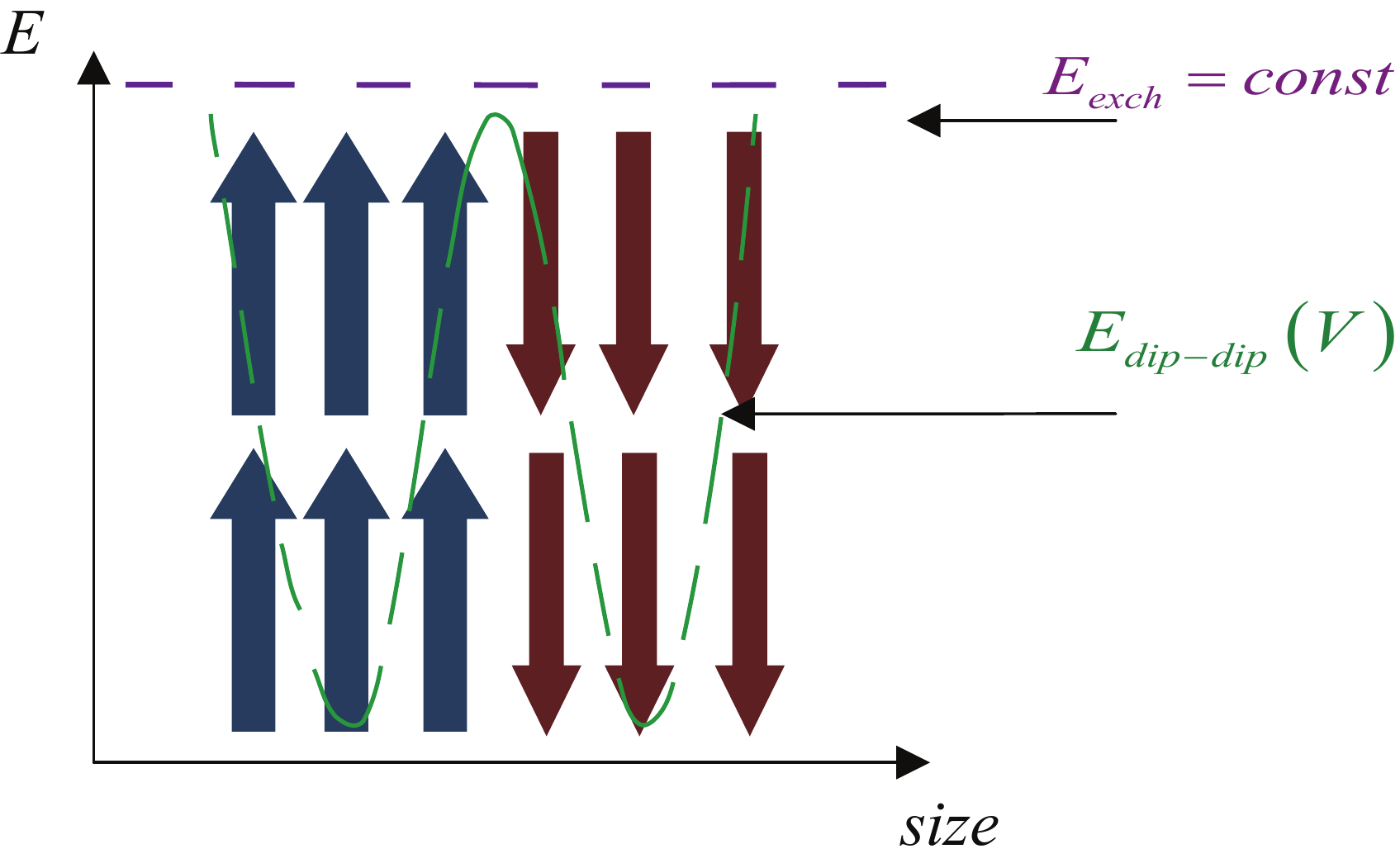}
	\caption{Diagram representation how \textit{exch} energy and \textit{dip-dip} energy behave with increasing of size.}
	\label{Fig1}
\end{figure}

Two competing interactions could be distinguished in ferromagnetic materials: an exchange interaction (\textit{exch}) which tends to orient magnetic moments in the same direction and by this magnetizes the system and dipole-dipole interaction (\textit{dip-dip}) which tends to orient magnetic moments antiparallel to each other and by this demagnetize the system. A relevant difference between these interactions is that \textit{exch} acts between nearest atoms and its energy is independent of a magnetic moment of a system. On the contrary, \textit{dip-dip} energy rises as magnetic moment increases. Increasing a material size a \textit{dip-dip} energy can overcome \textit{exch} energy. In this case, two and more domains structures become more favorable. It is schematically shown in figure \ref{Fig1}.

As \textit{exch} energy is much bigger than \textit{dip-dip} energy between nearest atoms magnetic moments inside a domain are firmly connected in the same direction. The magnetic shell where \textit{dip-dip} energy becomes equal to \textit{exch} energy could be accepted as a border of the domain and by this defining a size of a domain \cite{5}. So, after a domain was completed the next shell of magnetic moments will recline in the opposite direction, figure \ref{Fig1}. It should be mentioned that magnetic moments inside a domain are not strictly directed in the same direction. They recline under an angle to each other from shell to shell until the domain would not be finished. The transition layers behaviour is detailed described in the work of Landau-Lifshitz \cite{6}. Because of it, a magnetic moment of a domain is less than a sum of magnetic moments of atoms inside it.

On the assumption of foregoing we could consider domains like solitary magnetic particles. The separated quasiparticles we would call supermagneton (\textit{sm}) analogically to R.Harrison \cite{7}-\cite{9} where domains are also supposed to be a unit magnetic moments in the magnetization processes.

We know that \textit{exch} is compensated by \textit{dip-dip} between \textit{sm}s. It means that \textit{sm}s’ magnetic moments are exempt from \textit{exch}. So, the problem of ferromagnetic materials is brought to a problem of paramagnetic materials where \textit{sm}s play the role of magnetic moments.

Seeing that there are axes of easiest magnetization in ferromagnetic materials \cite{10} \textit{sm}s are distributed in a field of anisotropy according to the Boltzmann distribution. 

\begin{figure}[h!]
	\centering
	\includegraphics[width=0.4\textwidth]{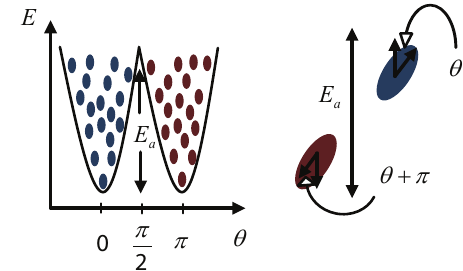}
	\caption{\textit{Sm}s separated by anisotropy barrier.}
	\label{Fig2}
\end{figure}

For simplicity, we would consider a case of uniaxial anisotropy. \textit{Sm}s with a positive projection on any selected direction along the anisotropy axis separated from \textit{sm}s with negative projection by anisotropy barrier, figure \ref{Fig2}.
\section{\label{sec:level1}III.	MAGNETIZATION PROCESS}
By applying magnetic field \textit{H} to a ferromagnetic material a magnetic field \textit{B} is induced inside. \textit{Sm}s with positive projection on the magnetic field direction obtain energy -$ \textit{m}^{+}\textit{B} $, and \textit{sm}s with negative projection obtain energy  $ \textit{m}^{-}\textit{B} $, where $ \textit{m}^{+} $  and  $ \textit{m}^{-} $ are magnetic moments of \textit{sm}s with positive and negative projection respectively.

Taking into account that  $ \textit{m}^{+} $ + $ \textit{m}^{-} $ = 2$ \textit{m} $, \textit{m} is magnetic moment of \textit{sm}s in zero field, we can conclude that ${m^-}B+{m^+}B= 2mB$. It means that domains could be effectively replaced by \textit{sm}s magnetic moment of which remains unchanged and equal to a magnetic moment of domain in nonmagnetized state.

\begin{figure}[h!]
		\centering
	\includegraphics[width=0.3\textwidth]{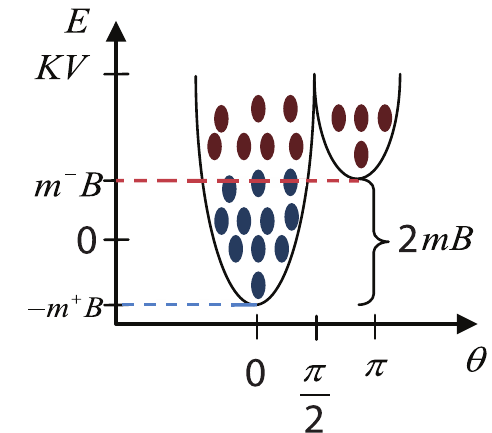}
	\caption{Shift of potential wells in the induced field \textit{B}.}
	\label{Fig3}
\end{figure}

So, potential wells shift on a value 2\textit{mB}, as shown on figure \ref{Fig3}.

As a result of the energetic shift magnetization of the sample is appeared.

To estimate the magnetization one needs to calculate a difference between magnetic moments with positive and negative projections. Here and in further positive and negative directions would be considered relative to the direction in which external magnetic field is applied.

Taking into account that at any field value distribution of \textit{sm}s is obey to the Boltzmann statistics we can calculate a difference between amount of \textit{sm}swith negatie and positive projections:
  \begin{equation}
  \begin{gathered}
  {{\text{N}}_1}-{{\text{N}}_2} =  \hfill \\
  \int\limits_{}^{} {g\left( {E + {E_a}} \right)} {e^{ - \frac{E}{{{k_B}T}}}}dE -  \hfill \\
  - \int\limits_{}^{} {g\left( {E - {E_a} - 2mB} \right)} {e^{ - \frac{E}{{{k_B}T}}}}dE =  \hfill \\
  = {e^{ - \frac{{{E_a}}}{{{k_B}T}}}} - {e^{ - \frac{{2mB - {E_a}}}{{{k_B}T}}}} \hfill \\ 
  \end{gathered}
  \label{eq1}
  \end{equation}
   \begin{equation}
   {{\text{N}}_1}+{{\text{N}}_2}={e^{-\frac{{{E_a}}}{{{k_B}T}}}}\left({1+{e^{-\frac{{2mB-2{E_a}}}{{{k_B}T}}}}}\right)
   \label{eq2}
   \end{equation}
   By the same way we get:
   \begin{equation}
   {{\text{N}}_1}+{{\text{N}}_2}={e^{-\frac{{{E_a}}}{{{k_B}T}}}}\left({1+{e^{-\frac{{2mB-2{E_a}}}{{{k_B}T}}}}}\right)
   \label{eq3}
   \end{equation}
   Dividing \ref{eq2} on \ref{eq3} we get:
   \begin{equation}
   \begin{gathered}
   {{\text{N}}_{\text{1}}}{\text{-}}{{\text{N}}_2}=\frac{{1-{e^{-\frac{{2\left({mB+{E_a}}\right)}}{{{k_B}T}}}}}}{{1+{e^{-\frac{{2\left({mB+{E_a}}\right)}}{{{k_B}T}}}}}}=\hfill \\
   =\left({{{\text{N}}_{\text{1}}}{\text{+}}{{\text{N}}_{\text{2}}}}\right)\tanh\left[{\frac{{2\left({mB-{E_a}}\right)}}{{{k_B}T}}}\right]\hfill \\
   \end{gathered}\
   \label{eq4}
   \end{equation}
   Multiplying  \ref{eq4} on \textit{m} and taking into account that by deffinition:
   \begin{equation}
   \left\{ {\begin{array}{*{20}{c}}
   	{{M_S} = m\left( {{{\text{N}}_{\text{1}}}{\text{ + }}{{\text{N}}_2}} \right)}  \\ 
   	{M = m\left( {{{\text{N}}_{\text{1}}}{\text{ - }}{{\text{N}}_{\text{2}}}} \right)}  \\ 
   	\end{array} } \right.\
   \label{eq5}
   \end{equation}
   finally we obtain:
   \begin{equation}
   \textit{M} = {M_S}\tanh \left[ {\frac{{2\left( {mB - {E_a}} \right)}}{{{k_B}T}}} \right]\
   \label{eq6}
   \end{equation}
   where $ \textit{M}_{S} $ is saturation magnetization and \textit{M}  is magnetization in field \textit{B}.
   
To explore magnetization process more thoroughly it is necessary to understand field distribution inside a material during magnetization. As ф direction of
external field is given we will concentrate on distribution of dipolar field.
    
The magnetic field which is induced by magnetic moment  in any point could be calculated in first approximation as \cite{12}:
\begin{equation}
\vec H = \frac{{3\hat n\left( {\vec m \cdot \hat n} \right) - \vec m}}{{{r^3}}}\
\label{eq7}
\end{equation}
where $ \vec n\ $ is a unit vector in the direction to the point, and \textit{r} is distance between magnetic moment and the point where the field is calculated.
	\begin{figure}[h!]
		\centering
		\includegraphics[width=0.3\textwidth]{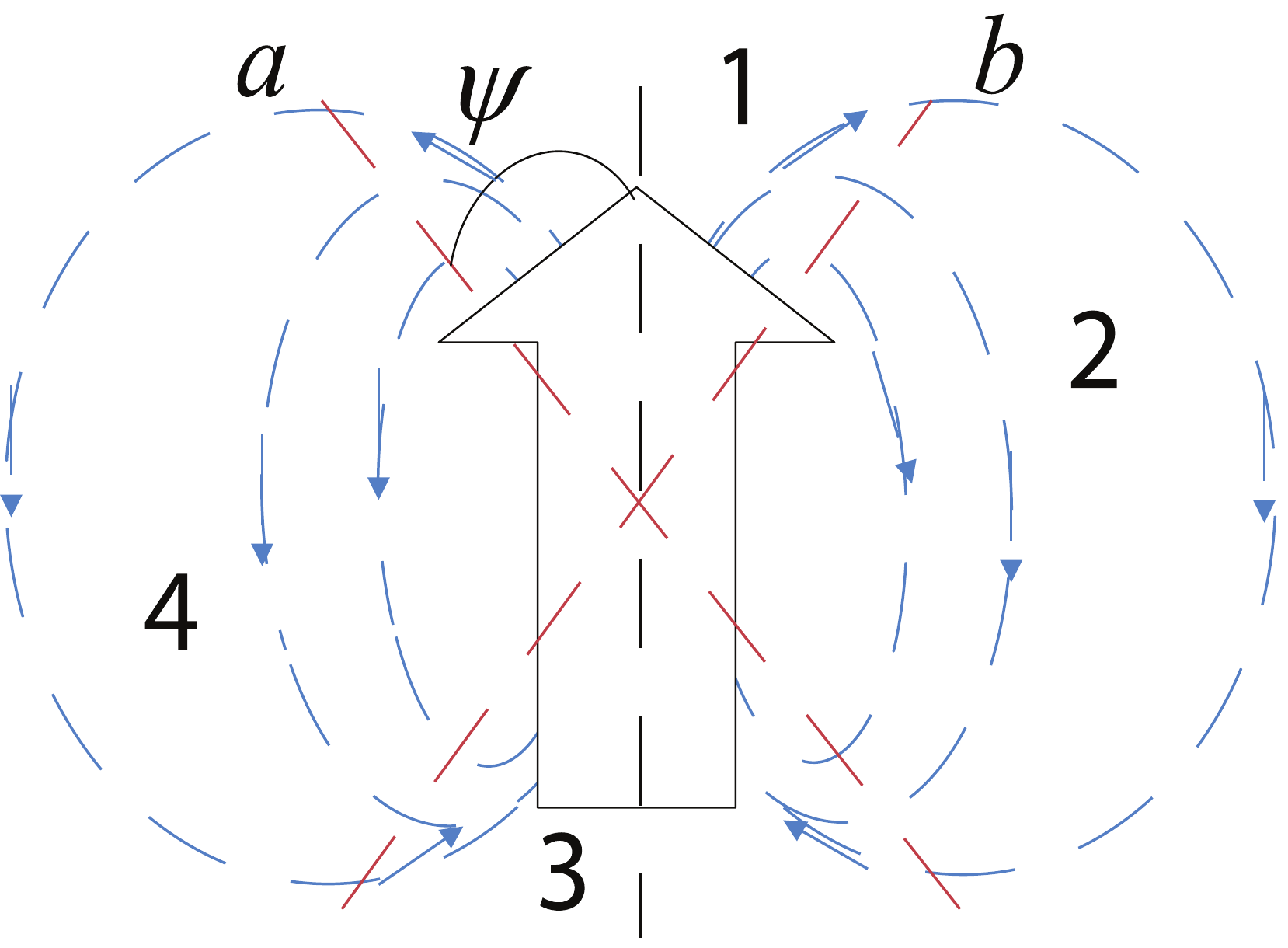}
		\caption{Dipolar field distribution around the magnetic moment. \textit{a} and \textit{b} are lines where magnetic field change their projection sing on magnetization direction. $ \psi \ $ is an angle between magnetization direction and lines \textit{a} and \textit{b}. 1, 3 are space regions where dipolar field has positive projection on magnetic moment direction and 2, 4 are space regions where dipolar field has negative projection on magnetic moment direction.}
		\label{Fig4}
	\end{figure}
	
It is seen from \ref{eq7} that \textit{H} has negative or positive projection on the direction of magnetic moment in different points.
It is not difficult to found points where \textit{H} changes its projection sign on direction of the magnetization from \ref{eq7}.
\begin{equation}
3\overset{\lower0.5em\hbox{$\smash{\scriptscriptstyle\frown}$}}{n} \left( {\vec M \cdot \overset{\lower0.5em\hbox{$\smash{\scriptscriptstyle\frown}$}}{n} } \right) - \vec M = 0\
\label{8}
\end{equation}
\begin{equation}
\left\{ {\begin{array}{*{20}{c}}
	{3{{\left( {\vec M\overset{\lower0.5em\hbox{$\smash{\scriptscriptstyle\frown}$}}{n} } \right)}^2} - \vec M\overset{\lower0.5em\hbox{$\smash{\scriptscriptstyle\frown}$}}{n}  = 0}  \\ 
	{\vec M\overset{\lower0.5em\hbox{$\smash{\scriptscriptstyle\frown}$}}{n}  = M\cos \left( \psi  \right)}  \\ 
	\end{array} } \right.\
\label{9}
\end{equation}
\begin{equation}
\cos \left( \psi  \right) = \frac{{\sqrt 3 }}{3}\
\label{10}
\end{equation}
\begin{equation}
\psi  \approx 55^\circ \
\label{11}
\end{equation}

These points belong to the lines \textit{a} and \textit{b} which decline under angles $ \psi \ $ to magnetization direction as shown on figure \ref{Fig4}. Lines \textit{a} and \textit{b} demarcate space around magnetic moment on four regions:  1, 2, 3, 4.

So, \textit{H}  has positive projection in any point which belongs to regions 1 and 3 with biggest value  when  or   and has negative projection in any pint of region 2 and 4 with biggest value $ \vec H = \frac{{2\vec m}}{{{r^3}}}\ $ when $ \psi  = 0^\circ \ $ or $ 180^\circ \ $. 

By this, dipolar field plays both magnetize (positive) and demagnetize (negative) roles. Due to positive influence of depolar field it is possible to magnetize bulk samples. 

Thus magnetic field inside a sample could be represented as:
\begin{equation}
\textit{B}=\textit{H}-\eta\frac{\textit{M}}{\textit{a}^3}\
\label{12}
\end{equation}
where $ \eta\frac{\textit{M}}{\textit{a}^3} $ is a dipolar term, $ \textit{a} $ is a distance between nearest domains domains (linear size of a $ \textit{sm} $), $ \eta $ is a coefficient which represents difference between positive and negative dipolar influences.

\begin{figure}[h!]
	\centering
	\includegraphics[width=0.3\textwidth]{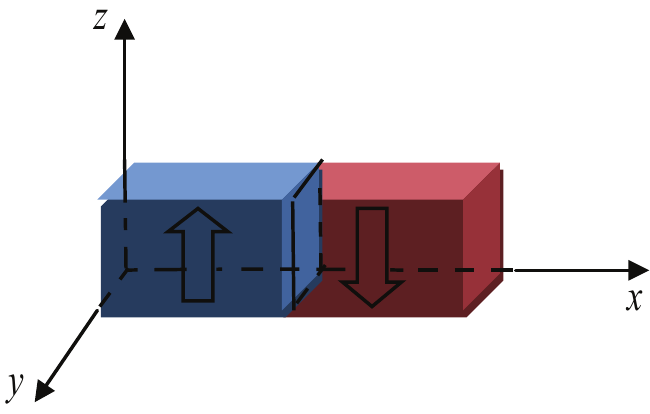}
	\caption{Two domain rod.}
	\label{Fig8}
\end{figure}

As an example, we will consider two-domain rod as shown on figure \ref{Fig8}. Let’s compare processes of magnetization of such a rod when the external magnetic field is directed along longitude (x – axis) and when the external magnetic field is directed along width (y – or z – axes).

\begin{figure}[h!]
	\centering
	\includegraphics[width=0.3\textwidth]{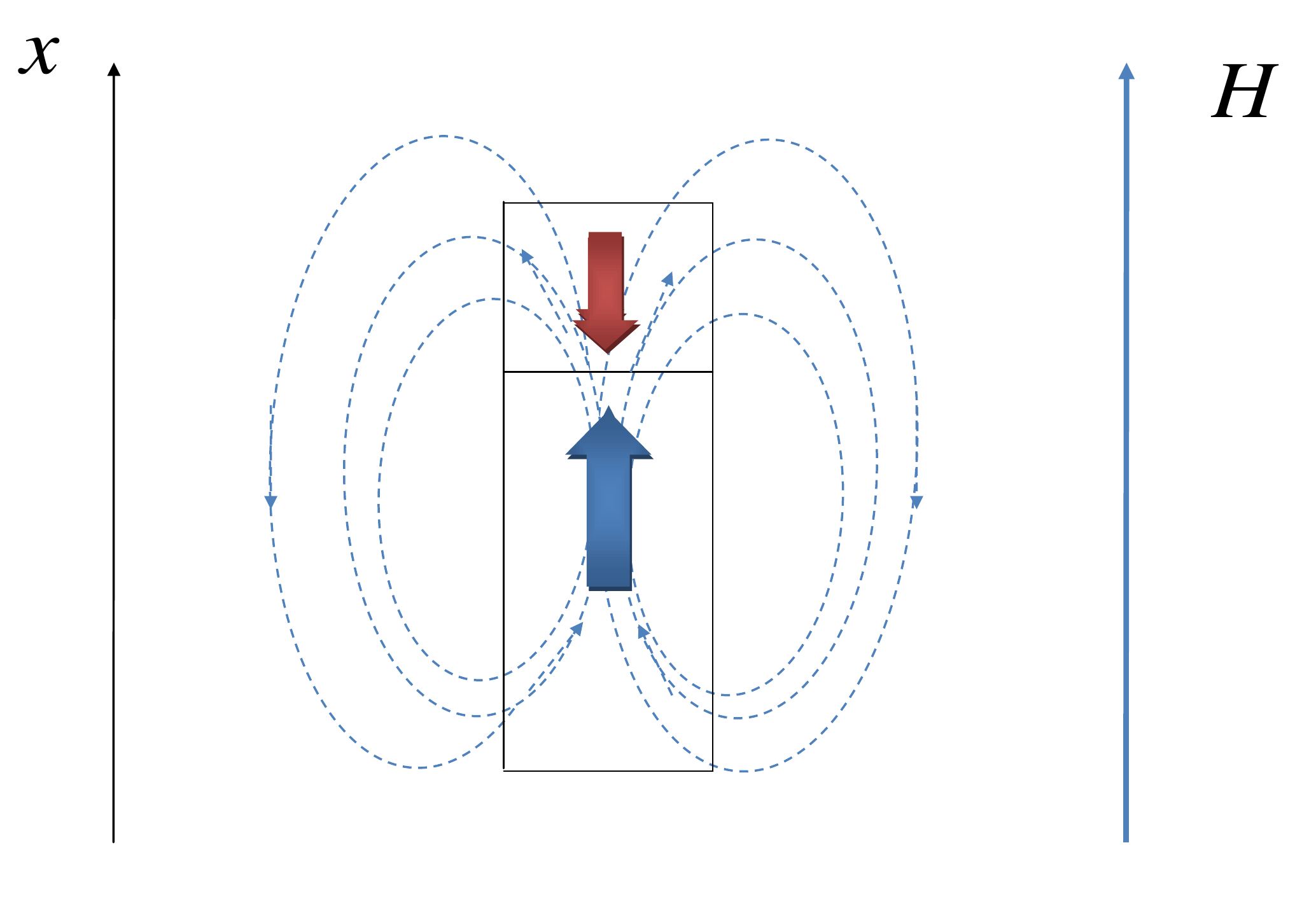}
	\caption{The rod is magnetized along longitude.}
	\label{Fig9}
\end{figure}

When the rod is magnetized along x axis the dipolar field of one domain will magnetize the second domain because it belongs to the region of the space where dipolar field has positive projection on magnetization direction. The diagram of the process is shown in figure \ref{Fig9}. By this dipolar field will help to magnetize the system. In this case s-shaped hysteresis loop would be observed \cite{7}-\cite{8}.
When the rod is magnetized along z axis one domain will direct the second domain in the opposite direction because it belongs to the region of the space where dipolar field has negative projection on magnetization direction as shown in figure \ref{Fig10}. In this case $ \eta  = 1 $.
\begin{figure}[h!]
	\centering
	\includegraphics[width=0.3\textwidth]{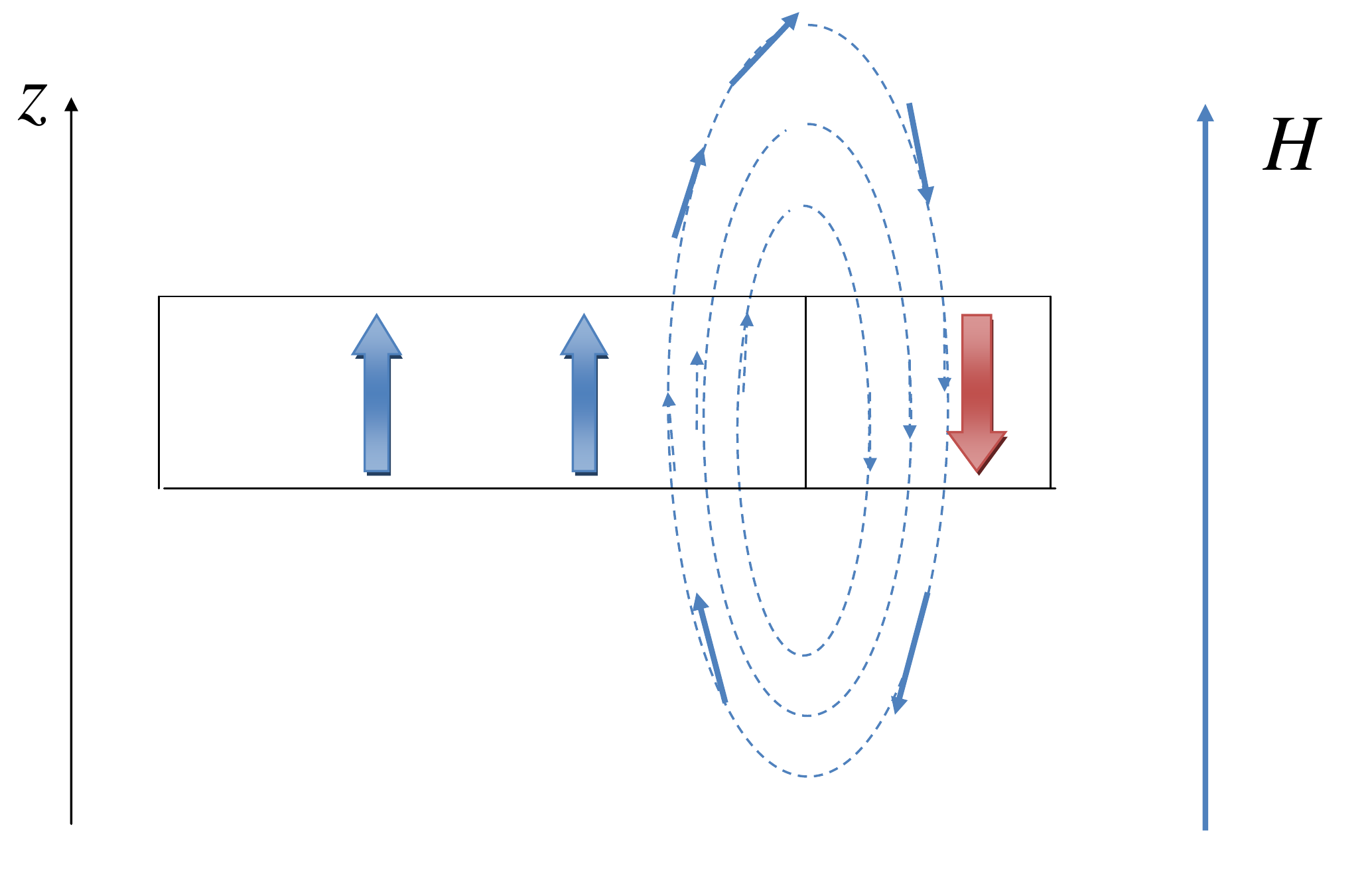}
	\caption{The rod is magnetized along width.}
	\label{Fig10}
\end{figure}
Because of all written above, the magnetic energy term in \ref{eq6} should be replaced by $\eta m{M_a}$, where $ \eta $ is a parameter which depend on difference between demagnetizing and magnetizing parts of \textit{dip-dip} influences, by this $ \eta $ depends on should depend on surface and surface – volume ratio. At each certain values of external magnetic field the certain distribution of field inside a material exists. This field distribution changes with external field, and by this $ \eta $ change with external field as well.
\begin{equation}
M = {M_S}\tanh \left[ {\frac{{mH - \eta \frac{m}{{{a^3}}}M + {E_a}\left( \gamma  \right)}}{{{k_B}T}}} \right]\
\label{eq13}
\end{equation}
As equation \ref{eq13} is trancendent it could be rewritten in more convenient form as done in \cite{7}-\cite{9}:
\begin{equation}
H = \frac{\eta }{{{a^3}}}M + \frac{{{k_B}T}}{{2m}}\ln \left( {\frac{{{M_S} - M}}{{{M_S} + M}}} \right) + \frac{{{E_a}\left( \gamma  \right)}}{m} \
\label{eq14}
\end{equation}
It should be noticed that the last term in \ref{eq14} is independent on the value of magnetic moment as both magnetic moment of the domains and anisotropy energy term are both volume dependant.

\section{\label{sec:level1}REMANENT MAGNETIZATION AND COERCIVE FORCE}
After the external magnetic field was abolished a \textit{dip-dip} tends to demagnetize the sample. Because of this \textit{sm}s would turn from positive projection to negative through anisotropy barrier. But not all \textit{sm}s can overcome anisotropy barrier and the part of them would remain with positive projection. So, after the external magnetic field was abolished ferromagnetic sample would steel remain in magnetize condition. This magnetization is called remanent magnetization. It is represented in figure \ref{11}. 
In order to get expression for remanent magnetization it is necessary to put \textit{H}=0 into \ref{eq13}:
\begin{equation}
{M_R} = {M_S}\tanh \left[ {\frac{{ - \eta \frac{m}{{{a^3}}}{M_R} + {E_a}\left( \gamma  \right)}}{{{k_B}T}}} \right]
\label{eq15}
\end{equation}
where $ \textit{M}_{R} $ is remanent magnetization.

As is known a coercive force is a magnetic field which should be applied to the sample to demagnetize it. So to get an expression of coercive force one need to put \textit{M}=0 into \ref{eq14}:
\begin{equation}
{H_C}\left( T \right) =  - \frac{{{E_a}\left( \gamma  \right)}}{m}
\label{eq16}
\end{equation}

It is seen that coercive force depends on the field direction (angle between field and anisotropy axis).
\begin{figure}[h!]
	\centering
	\includegraphics[width=0.3\textwidth]{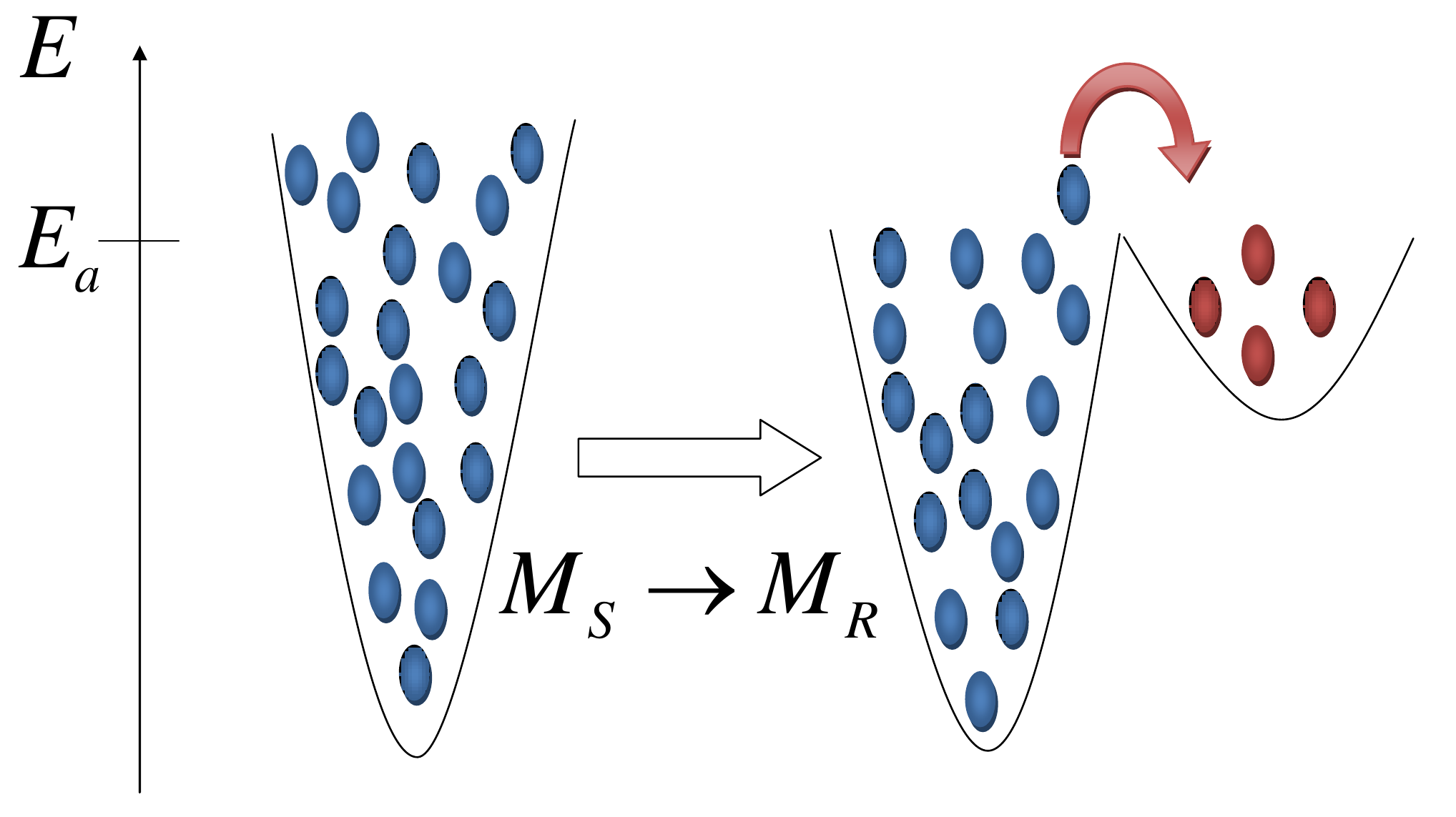}
	\caption{\textit{sm}s overcome anisotropy barrier when sample change its state from saturated magnetization to remanent magnetization.}
	\label{Fig10}
\end{figure}

\section{\label{sec:level1}MAGNETIC STIFFNESS}
The next parameter can characterize magnetic stiffness of ferromagnetic materials:
\begin{equation}
k = {e^{-2\frac{{\eta m{M_{{R_a}}} - {E_a}\left( \gamma  \right)}}{{{k_B}T}}}}
\label{eq17}
\end{equation}

Dividing \ref{eq2} on \ref{eq3} it is possible to show that $ k = \frac{{{N_2}}}{{{N_1}}} $ and represents the ratio of the amount of \textit{sm}s which overcame anisotropy barrier to the amount of \textit{sm}s which remain with positive projection after the external magnetic field was abolished, figure \ref{Fig11}.

Values of \textit{k} could change in the range from 0 to 1. The bigger the value of \textit{k} the softer magnetic material is and vice verse. For example, in case of strong anisotropy no of \textit{sm}s are able to overcome anisotropy barrier, it means that there are no \textit{sm}s with negative projection (\textit{k} = 0) and  – $ {M_R} = {M_S} $ rectangular-like hysteresis loop, figure \ref{12} a). In the case when all \textit{sm}s were able to overcome anisotropy barrier, the amount of \textit{sm}s with negative projection is equal to the amount of positive projection (\textit{k} = 1 ) and consequently $ {M_{\text{R}}} = 0 $, figure \ref{12} c).
\begin{figure}[h!]
	\centering
	\includegraphics[width=0.3\textwidth]{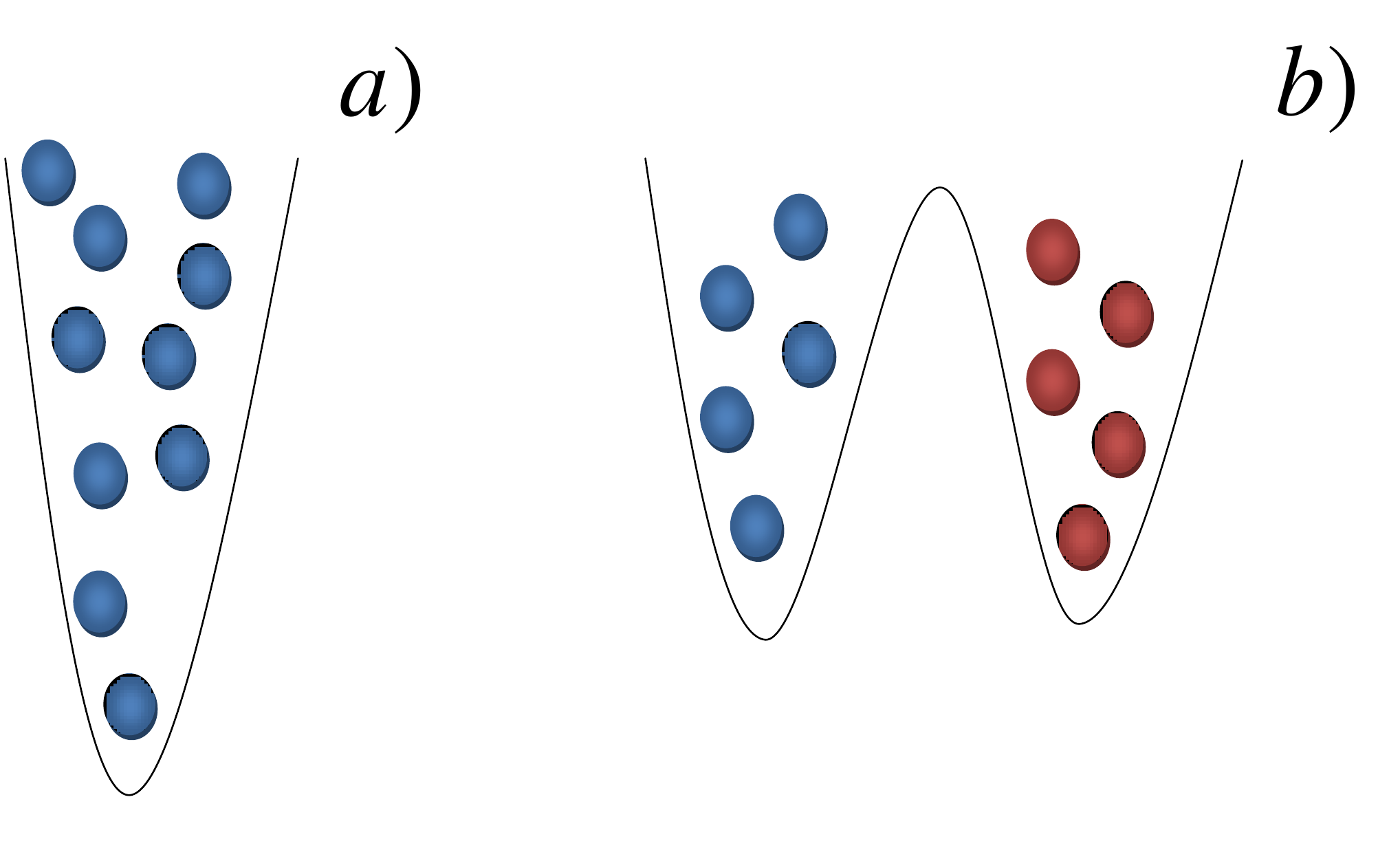}
	\caption{Three possible case of \textit{sm}s distribution in a state of remanent magnetization: a)$k = 0;$ b)$k = 1;$;  Cases a) and b) corresponds to a magneto stiff  material and magneto soft  materials correspondingly.}
	\label{11}
\end{figure}
\section{\label{sec:level1}DYNAMIC MAGNETIC SUSCEPTIBILITY}
A low which describes magnetization dependence on external magnetic field gives possibility to find out a law for magnetic susceptibility:
\begin{equation}
\begin{gathered}
\chi  = \frac{{dM}}{{dH}} = {M_S}\frac{d}{{dH}}\tanh \left[ {\frac{{m\left( {H - \eta {M_a}} \right) + {E_a}\left( \gamma  \right)}}{{{k_B}T}}} \right] =  \hfill \\
= \frac{{{M_S}m \cdot \left( {1 - \frac{{d\eta }}{{dH}}{M_a} - 4\pi \frac{\eta }{{{a^3}}}\chi } \right) + {M_S}{\raise0.7ex\hbox{${d{E_a}\left( \gamma  \right)}$} \!\mathord{\left/
				{\vphantom {{d{E_a}\left( \gamma  \right)} {dH}}}\right.\kern-\nulldelimiterspace}
			\!\lower0.7ex\hbox{${dH}$}}}}{{{k_B}T \cdot {\text{c}}{{\text{h}}^2}\left[ {\frac{{m\left( {H - \eta {M_a}} \right) + {E_a}\left( \gamma  \right)}}{{{k_B}T}}} \right]}} \hfill \\ 
\end{gathered}
\label{eq18}
\end{equation}
\begin{equation} 
\chi  = \frac{{{M_S}\left[ {m\left( {1 - \frac{{d\eta }}{{dH}}{M_a}} \right) + {\raise0.7ex\hbox{${d{E_a}\left( \gamma  \right)}$} \!\mathord{\left/
					{\vphantom {{d{E_a}\left( \gamma  \right)} {dH}}}\right.\kern-\nulldelimiterspace}
				\!\lower0.7ex\hbox{${dH}$}}} \right]}}{{{k_B}T{\text{c}}{{\text{h}}^2}\left[ {\frac{{m\left( {H - \eta {M_a}} \right) + {E_a}\left( \gamma  \right)}}{{{k_B}T}}} \right] + \frac{\eta }{{{a^3}}}m{M_S}}}
\label{eq19}
\end{equation}

It should be noted that $ \eta $ decreases with increasing of \textit{H} thus $ {\raise0.7ex\hbox{${d\eta }$} \!\mathord{\left/
		{\vphantom {{d\eta } {dH}}}\right.\kern-\nulldelimiterspace}
	\!\lower0.7ex\hbox{${dH}$}}M < 0 $.

This term is one of factors which are responsible for high value of magnetic susceptibility of ferromagnetic materials. 

It is seen that a magnetic susceptibility depends on the external magnetic field, the dipolar interaction, the temperature of the sample and the anisotropy energy, as it was expected.

In case of high temperature or small external magnetic field $ \left( {\tanh \left( x \right) \to x;\eta  \to 1} \right) $ from \ref{eq19} one can get:
\begin{equation}
\chi  = \frac{{{M_S}\left( {m + {\raise0.7ex\hbox{${d{E_a}\left( \gamma  \right)}$} \!\mathord{\left/
					{\vphantom {{d{E_a}\left( \gamma  \right)} {dH}}}\right.\kern-\nulldelimiterspace}
				\!\lower0.7ex\hbox{${dH}$}}} \right)}}{{{k_B}T + {\raise0.7ex\hbox{${m{M_S}}$} \!\mathord{\left/
				{\vphantom {{m{M_S}} {{a^3}}}}\right.\kern-\nulldelimiterspace}
			\!\lower0.7ex\hbox{${{a^3}}$}}}}
\label{eq20}
\end{equation}

\section{\label{sec:level1}CONCLUSION}
There was shown that a magnetic sample could be considered as a set of magnetic particles, called superparamagnetons (\textit{sm}s). Magnetic moment of \textit{sm} is equal to a magnetic moment of domain plus magnetic moment of the domain wall, before external magnetic field was applied. The stark difference of the \textit{sm} is that it's magnetic moment doesn't change during magnetisation precess and \textit{sm}s are free from exchange interaction what gives a possibility to apply the Boltzmann statistics for them as it done for paramagnetic samples. 

Based on this assumption there was derived the analytical excretion to describe a dependence of magnetization of a ferromagnetic material on an internal \ref{eq6}.

Considering dipolar field distribution inside a material an analytical expression of magnetization dependence on an external magnetic field \ref{eq13} is also deduced.

It is important that all energies that take place in the process of magnetization are included additively. It means that additional energies like energies on domain walls pinning could be easily added in the formula in  \ref{eq13}. 

There were derived expressions for a remanent magnetization \ref{eq15} and a coercive force \ref{eq16} as special points on \textit{M}(\textit{H}) curve. It is seen that temperature dependence of remanent magnetization bears exponential character and depend on \textit{dip-dip} in a sample and on an anisotropy barrier. It is also seen how coercive force depend on anisotropy barrier and a magnetic moment of a domain.

A new parameter which characterizes a magnetic stiffness of a material and its temperature dependence is introduced \ref{eq17}. 

From \textit{M}(\textit{H}) function there was derived an expression for a magnetic susceptibility \ref{eq19}. It is shown that in extremal cases, like high temperature or low field, magnetization depends on field linearly, and magnetic susceptibility is field independent like in the paramagnetic case. It is essential to note that expressions \ref{eq19} – \ref{eq20} are applicable at all temperature regions.


\begin{thebibliography}{13}
\bibliographystyle{IEEEtran}
\bibitem{1} 
D.C Jiles, X. Fang, W. Zhang, 
\textit{Handbook of Advanced Magnetic Materials}, (2006).
\bibitem{2} 
F.Liorzou, B. Phelps, and D.L. Atherton, 
\textit{Macroscopic models of magnetization},IEEE transactions on magnetics, vol.36, No.2., (2000).
\bibitem{3} 
Sergey E. Zirka, Yuriy L. Moroz, Robert G. Harrison, and Krzysztof Chwastek, \textit{On the physical aspects of the Jiles-Atherton hysteresis models}, J. Appl. Phys. 112, (2012).
\bibitem{4}
D.C. Jiles and D.L. Atherton, "Theory of ferromagnetic hysteresis", J. Magn. Magn. Mater. 61, 48, (1986).
\bibitem{5}
J. Frankel and J. Dorfman; "Spontaneous and Induced Magnetisation in Ferromagnetic Bodies",  (1930).
\bibitem{6}
L. Landau, E. Lifshits, "On the theory of the dispersion of magnetic permeabillity in ferromagnetic bodies", Phys. Zeitsch. der Sow. 8, pp. 153-169, (1935).
\bibitem{7}
R.G. Harrison, "Physical model of spin ferromagnetism", (2003).
\bibitem{8}
R.G. Harrison, "Variable-Domain-Size theory of spin ferromagnetism",   (2004).
\bibitem{9}
R.G. Harrison, "Physical theory of ferromagnetic first-order return curves", IEEE TRANSITION ON MAGNETICS, Vol. 45 NO, 4, (2009).
\bibitem{10}
L. D. Landau and E. M. Lifshitz, Course of Theoretical Physics, Vol. 8, (2005).
\bibitem{11}
C. Kittel, "Theory of the structure of ferromagnetic domains in films and small particles", Phys.Rev. 	vol.70, NO. 11 and 12, (1946).
\bibitem{12}
L. D. Landau and E. M. Lifshitz, Course of Theoretical Physics, Vol. 2: The Classical Theory of Fields (Nauka, Moscow, 1988; Pergamon, Oxford, 1975).
\bibitem{13}
Alberto P. Gimaraes, Principles of Nanomagnetism, Springer, 2009.
\end{thebibliography}
\end{document}